\documentclass[10pt,conference,a4paper]{IEEEtran}
\usepackage{cite}

\ifCLASSINFOpdf
   \usepackage[pdftex]{graphicx}
\else
\fi
\usepackage{amsmath}
\usepackage{algorithmic}

\usepackage{array}
\usepackage{amsfonts} 
\usepackage{multicol}
\usepackage{amssymb}
\usepackage{cases}
\usepackage{overpic}
\usepackage{caption}
\usepackage{subcaption}
\usepackage{multirow}
\usepackage{tabularx}
\usepackage{enumitem}
\usepackage[ruled]{algorithm2e}
\usepackage[colorlinks]{hyperref}

\def\bx{{\bf x}}
\def\by{{\bf y}}
\def\bz{{\bf z}}

\def\0{{\bf 0}}
\def\1{{\bf 1}}

\def\bH{{\bf H}}

\def\bX{{\bf X}}
\def\bY{{\bf Y}}

\def\mbR{{\mathbb R}}

\DeclareMathOperator{\prox}{Prox}

\def\etal{\emph{et al. }}
\def\ie{\emph{i.e. }}

\begin{document}
%
\title{Deep Iterative Residual Convolutional Network for Single Image Super-Resolution}

\author{\IEEEauthorblockN{Rao M. Umer, G. L. Foresti, Senior member, IEEE, C. Micheloni, Member, IEEE}
\IEEEauthorblockA{University of Udine, Italy.}}


%


\maketitle

\begin{abstract}
Deep convolutional neural networks (CNNs) have recently achieved great success for single image super-resolution (SISR) task due to their powerful feature representation capabilities. The most recent deep learning based SISR methods focus on  designing deeper / wider models to learn the non-linear mapping between low-resolution (LR) inputs and high-resolution (HR) outputs. These existing SR methods do not take into account the image observation (physical) model and thus require a large number of network's trainable parameters with a great volume of training data. To address these issues, we propose a deep Iterative Super-Resolution Residual Convolutional Network (ISRResCNet) that exploits the powerful image regularization and large-scale optimization techniques by training the deep network in an iterative manner with a residual learning approach. Extensive experimental results on various super-resolution benchmarks demonstrate that our method with a few trainable parameters improves the results for different scaling factors in comparison with the state-of-art methods.
\end{abstract}


%
\IEEEpeerreviewmaketitle

\section{Introduction}
\label{sec:intro}
The goal of the single image super-resolution (SISR) is to recover the high-resolution (HR) image from its low-resolution (LR) counterpart. SISR problem is a fundamental low-level vision and image processing problem with various practical applications in  satellite imaging, medical imaging, astronomy, microscopy, seismology, remote sensing, surveillance, biometric, image compression, etc. In the last decade, most of the photos are taken using built-in smartphones cameras, where the resulting LR image is inevitable and undesirable due to their physical limitations. It is of great interest to restore sharp HR images because some captured moments are difficult to reproduce. On the other hand, we are also interested to design low cost (limited memory and cpu power) camera devices, where the deployment of our deep network would be possible in practice. Both are the ultimate goals to the end users. 

Usually, SISR is described as a linear forward observation model by the following image degradation process:
\begin{equation}
    \by = \bH \Tilde{\bx} + \eta,
    \label{eq:degradation_model}
\end{equation}
where $\by \in \mbR^{N/s^2}$ is an observed LR image (here $N=m\times n$ is typically the total number of pixels in an image), $\bH \in \mbR^{N/s\times N/s}$ is a \emph{down-sampling operator} (usually a bicubic, circulant matrix) that resizes an HR image $\Tilde{\bx} \in \mbR^{N}$ by a scaling factor $s$ and $\eta$ is considered as an additive white Gaussian noise with standard deviation $\sigma$. However, in real-world settings, $\eta$ also accounts for all possible errors during the image acquisition process that include inherent sensor noise, stochastic noise, compression artifacts, and the possible mismatch between the forward observation model and the camera device. The operator $\bH$ is usually ill-conditioned or singular due to the presence of unknown noise ($\eta$) that makes the SISR of a highly ill-posed nature of inverse problems. Since, due to ill-posed nature, there are many possible solutions, regularization is required to select the most plausible ones.

Generally, SISR methods can be classified into three main categories, \ie interpolation-based methods, model-based optimization methods, and discriminative learning methods. Interpolation-based methods \ie nearest-neighbor, bilinear, and bicubic interpolators are efficient and simple, but have very limited reconstruction image quality. Model-based optimization~\cite{Dong2013NonlocallyCS} methods have powerful image priors to reconstruct high-quality clean images, but require hundreds of iterations to achieve acceptable performance, thus making these methods computationally expensive. Model-based optimization~\cite{Zhang2017LearningDC} methods with the integration of deep CNNs priors can improve efficiency, but due to hand-crafted parameters, they are not suitable for end-to-end deep learning methods. On the other hand, discriminative learning~\cite{kim2016vdsrcvpr, Lim2017edsrcvprw,Umer_2020_CVPR_Workshops} methods have attracted significant attentions due to their effectiveness and efficiency for SISR performance by using deep CNNs. Our work is inspired by discriminative and residual learning approaches with powerful image priors and large-scale optimization schemes in an iterative manner for an end-to-end deep CNNs to solve SISR problem. 
\begin{figure*}[htbp!]
\centering
\includegraphics[width=11.0cm]{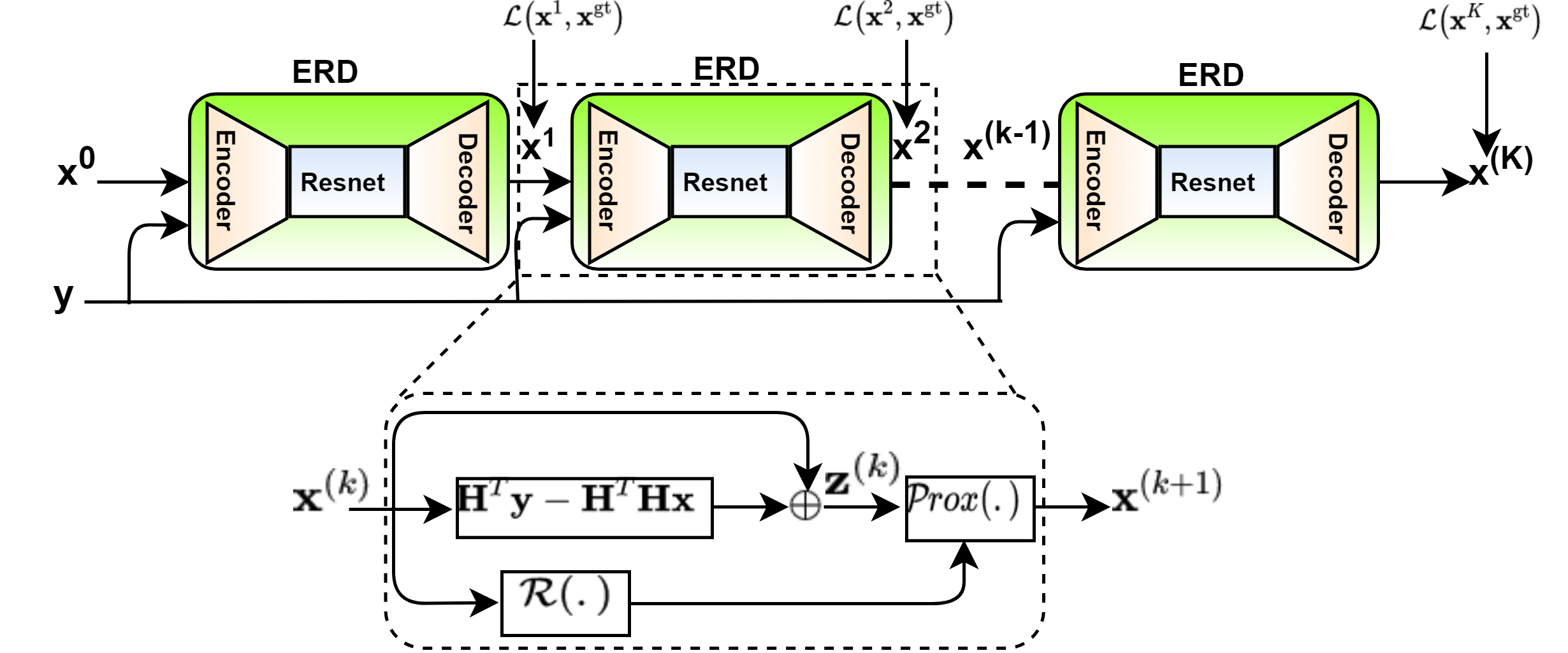}
\caption{The visualization of our proposed iterative SISR approach as described in Algorithm~\ref{alg:alg_sisr}. Given an LR image ($\by$) and an initial estimate ($\bx^0$), each network's stage \emph{ERD} (Encoder-Resnet-Decoder) produces a new estimate $\bx^{(k+1)}$ from the previous step estimate $\bx^{(k)}$. A single optimizer is used for all network stages with shared structures and parameters by $K$ steps.}
\label{fig:iter_mm_scheme}
\vspace{-0.1cm}
\end{figure*}
The visualization of our proposed iterative SISR approach is shown in Figure \ref{fig:iter_mm_scheme}, where the LR input ($\by$) is given to the network and then the network reconstructs the SR output. A single optimizer is used for all network stages with shared structures and parameters. Our contributions in this paper are three-fold as follows:
\begin{enumerate}
\item We propose an end-to-end deep iterative Residual CNNs for image super-resolution. In contrast to the existing deep SISR networks, our proposed method strictly follows the image observation (physical) model (refers to Eq.~\eqref{eq:degradation_model}), and thus it is able to achieve better reconstruction results even with few network's trainable parameters (refers to Table~\ref{tab:ablation_table}).
\item A deep SISR network is proposed to solve image super-resolution in an iterative manner by minimizing the discriminative loss function with a residual learning approach.
\item The proposed ISRResCNet is inspired by powerful image regularization and large-scale optimization techniques that have been successfully used to solve general inverse problems in the past.
\end{enumerate}

\section{Related Works}
\label{sec:related_work}
Recently, numerous works have been addressed the task of SISR that are based on deep CNNs for their powerful feature representation capabilities. A preliminary CNN-based method to solve SISR is a super-resolution convolutional network with three layers (SRCNN)~\cite{dong2014srcnneccv}. Kim~\etal\cite{kim2016vdsrcvpr} proposed a very deep SR (VDSR) network with residual learning approach. Lim \etal\cite{Lim2017edsrcvprw} proposed an enhanced deep SR (EDSR) network by taking advantage of the residual learning. Zeng~\etal\cite{zeng2018s3csr} proposed an S$_3$cSR method (conventional sparse coding) that learns HR/LR dictionaries exploiting the iterative sparse coding. Jiang~\etal\cite{jiang2018risr} proposed a Recursive Inception (RISR) network to adopt the inception-like structure to extract HR/LR features. Liu~\etal\cite{liu2018atsr} proposed an attention-based approach for SISR problem. Yaoman \etal\cite{Li2019srfbncvpr} proposed a feedback network (SRFBN) based on feedback connections and recurrent neural network like structure. Zhang \etal\cite{zhang2019deep} proposed a deep plug-and-play Super-Resolution method for arbitrary blur kernels by following the multiple degradation.  In \cite{srwdnet}, the authors proposed SRWDNet to solve the joint deblurring and super-resolution task by following the realistic degradation. The above methods are deep or wide CNN networks to learn non-linear mapping from LR to HR with a large number of training samples, while neglecting the image acquisition process. However, our approach takes into account the physical image observation process by greatly increase its applicability. 

\section{Proposed Method}
\subsection{Problem Formulation}
\label{sec:prob_form}
By referencing to equation~\eqref{eq:degradation_model}, the recovery of $\bx$ from $\by$ mostly relies on the variational approach for combining the observation and prior knowledge, and is given as the following objective function:
\begin{equation}
    \mathbf{J}(\bx) = \underset{\mathbf{x}}{\arg \min }~\frac{1}{2}\|\by - \bH \bx\|_2^{2}+\lambda \mathcal{R}(\bx),
    \label{eq:eq1}
\end{equation}
where $\frac{1}{2}\|\by-\mathbf{H} \bx\|_2^2$ is the data fidelity (also known as log-likelihood) term that measures the proximity of the solution to the observations, $\mathcal{R}(\bx)$ is the regularization term that is associated with image priors, and $\lambda$ is the trade-off parameter that governs the compromise between the data fidelity and the regularizer term. Interestingly, the variational approach has a direct link to the Bayesian approach and the derived solutions can be described by either as penalized maximum likelihood or as maximum a posteriori (MAP) estimates~\cite{bertero1998map1,figueiredo2007map2}. Thanks to the recent advances of deep learning, the regularizer (\ie $\mathcal{R}(\bx)$) is employed by deep convolutional neural networks (ConvNets) that have powerful image priors capabilities. 

\begin{figure*}[htbp!]
\centering
\includegraphics[width=12cm]{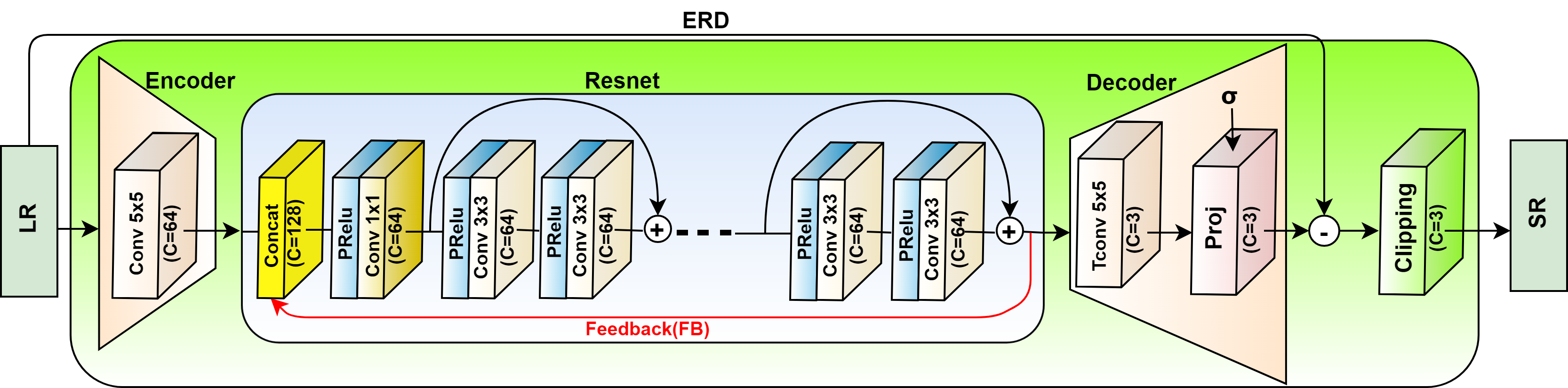}
\caption{The architecture of ERD (Encoder-Resnet-Decoder) blocks used in the proposed ISRResCNet.}
\label{fig:ERD_arch}
\end{figure*}

\subsection{Objective Function Minimization Strategy}
\label{subsec:obj_func_min}
Besides the proper selection of the regularizer and formulation of the objective function, another important aspect of the variational approach is the minimization strategy that will be used to get the required solution. In the literature, there are several modern convex-optimization schemes for large-scale machine learning problems, such as Split-Bregman~\cite{split_bregman}, HQS method~\cite{HQS}, ADMM ~\cite{boyd_admm}, Primal-dual algorithms~\cite{Chambolle2011Primaldual}, etc. In our work, we solve the under study problem~\eqref{eq:eq1} by using the Majorization-Minimization (MM) framework~\cite{hunter2004tutorial} because $\mathbf{J}(\bx)$ is too complicated to manipulate (\ie convex function but possibly non-differentiable). In MM~\cite{hunter2004tutorial,figueiredo2007majorization,lefkimmiatis2011hessian} approach, an iterative algorithm for solving the minimization problem
\begin{equation}
    \hat{\bx} = \underset{\mathbf{x}}{\arg \min }~\mathbf{J}(\bx),
    \label{eq:eq2}
\end{equation}
takes the form
\begin{equation} 
    \bx^{(k+1)} = \underset{\mathbf{x}}{\arg \min }~ \mathbf{Q}(\bx;{\bx}^{(k)}),
    \label{eq:eq3}
\end{equation}
where $\mathbf{Q}(\bx;{\bx}^{(k)})$ is the majorizer of the function $\mathbf{J}(\bx)$ at a fixed point ${\bx}^{(k)}$ by satisfying the following two conditions:
\begin{equation}
    \mathbf{Q}(\bx;{\bx}^{(k)})>\mathbf{J}(\bx), ~\forall \bx \ne \bx^{(k)} \quad\\ ~\mbox{and}~\quad \\ 
    \mathbf{Q}(\bx^{(k)};\bx^{(k)}) = \mathbf{J}(\bx^{(k)}).
    \label{eq:eq4}
\end{equation}
Here, we want to upper-bound the $\mathbf{J}(\bx)$ by a suitable majorizer $\mathbf{Q}(\bx;{\bx}^{(k)})$, and instead of minimizing the actual objective function \eqref{eq:eq2} due to its complexity, we minimize the majorizer $\mathbf{Q}(.)$ to produce the next estimate $\bx^{(k+1)}$. By satisfying the properties of the majorizer given in \eqref{eq:eq4}, iteratively minimizing $\mathbf{Q}(.;{\bx}^{(k)})$ also decreases the actual objective function $\mathbf{J}(.)$ \cite{hunter2004tutorial}. Thus, we can write a quadratic majorizer for the complete objective function~\eqref{eq:eq1} as the following form:  
\begin{equation}
    \mathbf{Q}(\bx;{\bx}^{(k)}) = \underset{\mathbf{x}}{\arg \min }~\frac{1}{2}\|\by - \bH \bx\|_2^{2}+\lambda \mathbf{Q}_\mathcal{R}(\bx;{\bx}^{(k)}),
    \label{eq:eq5}
\end{equation}
To start an initial estimate $\bx_0$, we have:
\begin{equation}
    \mathbf{Q}_\mathcal{R}(\bx;\bx_0) = \frac{1}{2} (\bx - \bx_0)^T [\alpha \mathbf{I} - \bH^T \bH ] (\bx - \bx_0),
    \label{eq:eq6}
\end{equation}
where $\mathbf{Q}_\mathcal{R}(.)$ is a distance function between $\bx$ and $\bx_0$. In order to get a valid majorizer $\mathbf{Q}_\mathcal{R}(.)$, we need to satisfy two conditions in \eqref{eq:eq4} as $\mathbf{Q}_\mathcal{R}(\bx;\bx_0)>0$, $\forall \bx \ne \bx_0$ and $\mathbf{Q}_\mathcal{R}(\bx;\bx_0)=0$. This suggests that $\alpha \mathbf{I} - \bH^T \bH $ must be a positive definite matrix, which only holds if $\alpha > \|\bH^T \bH\|_2$. The parameter $\alpha$ depends upon the largest eigenvalue of $\bH^T \bH$, but, in most image restoration cases~\cite{figueiredo2007majorization} such as inpainting, deblurring, demosaicking\cite{kokkinos2019iterative}, and super-resolution, it approximately equals to one ($\alpha\approx1$). Based on the above discussion, we can write the overall majorizer as:
\begin{equation}
    \mathbf{Q}(\bx;\bx_0) = \frac{1}{2/\alpha}\|\bx - \bz\|_2^{2}+\lambda \mathcal{R}(\bx) + const.,
    \label{eq:eq8}
\end{equation}
where $\bz = \bx_0 + \frac{1}{\alpha}\bH^T(\by - \bH\bx_0) $, and the \emph{constant} does not depend on $\bx$ and thus it is irrelevant to the optimization task. \\
Finally, we proceed with the MM optimization scheme to iteratively minimize the quadratic majorizer function $\mathbf{Q}(.)$ by the following formulation as:
\begin{equation}
\begin{split}
\hat{\bx}^{(k)} & = \underset{\mathbf{x}}{\arg \min }~\mathbf{Q}(\bx;\bx^k) \\
 & = \underset{\mathbf{x}}{\arg \min }~\frac{1}{2}\|\by - \bH \bx\|_2^{2}+\lambda \mathbf{Q}_\mathcal{R}(\bx;\bx^k) \\
 & = \underset{\mathbf{x}}{\arg \min }~\frac{1}{2/\alpha}\|\bx - \bz^k\|_2^{2}+\lambda \mathcal{R}(\bx) \\
 & = \prox_{(\lambda/\alpha)\mathcal{R}(.)}(\bz^k)
\end{split}
\label{eq:eq9}
\end{equation}
where $\bz^k = \bz^k + \bH^T(\by - \bH\bz^k) $ and $\prox_{(.)}$ is the proximal operator~\cite{ParikhPGM}, which is defined as: 
\begin{equation}
     \mathbf{P}_\mathbb{C}(\bz) = \arg\underset{\bx \in \mathbb{C}}{\min} ~\frac{1}{2\sigma^2}\|\bx - \bz\|_2^2 + \frac{\lambda}{\alpha}\mathcal{R}(\bx).
    \label{eq:eq10}
\end{equation} 
It can be noted that the above Eq.~\eqref{eq:eq10} is treated as the objective function of a denoising problem, where $\bz$ is the noisy observation with noise level $\sigma$. In this way, we heavily rely on employing a deep denoising neural network to get the required estimate $\hat{\bx}^{(k)}$ by unrolling the MM scheme as $K$ finite steps. Another thing to notice, in Eq.~\eqref{eq:eq9}, is that we decouple the degradation operator $\bH$ from $\bx$ and now we need to tackle it with a less complex denoising problem. However, obtaining the resulting solution $\hat{\bx}^{(k)}$ from~\eqref{eq:eq9} can be computationally expensive since it demands $K$ times the parameters of the employed denoiser and can exhibit the slow convergence\cite{beck2009fast,kokkinos2018deep}. To avoid this hurdles, we adopt the similar strategy as done in\cite{kokkinos2019iterative}, where the trainable extrapolation weights $\mathbf{w^{(k)}}$ are learnt directly from the training data instead of the fixed ones~\cite{li2015accelerated}. Moreover, the convergence of our proposed method is sped up by adopting the continuation strategy\cite{lin2014adaptive}. Our overall proposed method is shown in Fig.~\ref{fig:iter_mm_scheme} and also described in the Algorithm \ref{alg:alg_sisr}, where the input settings, initialization, extrapolation steps, and proximal steps are defined. Our proposed Algorithm \ref{alg:alg_sisr} has a close connection to other proximal algorithms such as ISTA~\cite{daubechies2004ista} and FISTA~\cite{beck2009fista} that require the exact form of the employed regularizer such as Total Variation / Hessian Schatten-norm~\cite{lefkimmiatis2011hessian}. However, in our case, the regularizer is learned implicitly from the training data (\ie non-convex form), and therefore our algorithm acts as an inexact form of proximal gradient descent steps.\\   
\begin{algorithm}[t]
 \SetAlgoCaptionSeparator{\unskip:}
 \SetKwInOut{Input}{Input}
 \SetKwInOut{Output}{Output}
 \Input{ $\by$: LR input, $\bH$: Down-sampling operator, $\bH^T$: Up-sampling operator, $K$: iterative steps, $\mathbf{w}\in\mbR^K$: extrapolation weights, $\sigma$: estimated noise, $\lambda,\alpha$: projection parameters}%
 \textbf{Initialization:} $\bx^{(0)}= \bH^T\by$, $\bH^T$ : Bilinear kernel\;
                        $\bz^{(1)} = \bx^{(0)} + \bH^T(\by - \bH\bx^{(0)})$\;
  \For{$k\gets1$ \KwTo $K$}{
    Extrapolation step: $ \bz^{(k+1)} = \bx^{(k)} + \mathbf{w^{(k)}} (\bx^{(k)} - \bx^{(k-1)}) $\;
    Proximal step (ERD-block): $\hat{\bx}^{(k)}=\prox_{(\lambda/\alpha)\mathcal{R}(.)}(\bz^k + \bH^T(\by - \bH\bz^k))$\;
 }
 \Output{$\bx^K$ : SR output}
 \caption{The proposed SISR iterative approach. The ERD structure and parameters are shared across all iterative steps.}
 \label{alg:alg_sisr}
\end{algorithm}

\subsection{Network Architecture}
\label{sec:network_arch}
The proposed network architecture for super-resolution is shown in Fig.~\ref{fig:iter_mm_scheme}.  Given an LR image ($\by$) and an initial estimate ($\bx^0$), each network's stage \emph{ERD} (Encoder-Resnet-Decoder) produces a new estimate $\bx^{(k+1)}$ from the previous step estimate $\bx^{(k)}$. The Algorithm~\ref{alg:alg_sisr} describes the inputs, initial conditions, and desired updates for each network stage. The ERD structure and parameters are shared across all iterative steps. Finally, a single optimizer is used to minimize the $\ell_1$-Loss between the estimated latent SR image ($\bx^{(k)}$) and ground-truth (GT) ($\bx^{(gt)}$) after k-steps as:\\
\begin{equation}
    \arg\underset{\Theta}{\min}~\mathcal{L}(\mathbf{\Theta}) = \frac{1}{2} \sum \limits _{n = 1}^{N} \|\bx_n^{k} - \bx_{n}^{gt}\|_1
    \label{eq:l1loss}
\end{equation}
where $N$ is the mini-batch size and $\mathbf{\Theta}$ are the trainable parameters of our network. Fig.~\ref{fig:ERD_arch} shows the \emph{ERD} block used in the network. In ERD network, both \emph{Encoder (Conv)} and \emph{Decoder (TConv)} layers have $64$ feature maps of $5\times5$ kernel size with $C \times H\times W$ tensors, where $C$ is the number of channels of the input image. \emph{Resnet} consists of $5$ residual blocks with two Pre-activation \emph{Conv} layers, each of $64$ feature maps with kernels support $3\times3$, and the pre-activation is the parametrized rectified linear unit (PReLU)\cite{He2015DelvingDI} with $64$ out feature channels. The \emph{Resnet} also contains the Feedback (FB) path after $5$ resblocks with an initial concatenation pre-activation \emph{Conv} layer by $1\times1$ kernel support that maps 128 features channels to 64 to feed into resblocks. The trainable projection layer~\cite{Lefkimmiatis2018UDNet} inside the \emph{Decoder} computes the proximal map  for Eq.~\eqref{eq:eq10} with given noise standard deviation $\sigma$ and handle the data fidelity and prior terms. The noise realization is estimated in the intermediate \emph{Resnet} that is sandwiched between the \emph{Encoder} and \emph{Decoder}. The estimated residual image after \emph{Decoder} is subtracted from the LR input image. Finally, the clipping layer incorporates our prior knowledge about the valid range of image intensities and enforces the pixel values of the reconstructed image to lie in the range $[0, 255]$. Reflection padding is also used before all \emph{Conv} layers to ensure slowly-varying changes at the boundaries of the input images. Our ERD structure can also be described as the generalization of one stage TNRD~\cite{chen2017tnrdtpami} and UDNet~\cite{Lefkimmiatis2018UDNet} that have good reconstruction performance for image denoising problem.

\subsection{Network Training via TBPTT}
Due to the iterative nature of our SISR approach, the network parameters are updated using back-propagation through time (BPTT) algorithm by unrolling $K$ steps to train the network, which is previously used in recurrent neural networks training such as LSTMs. However, it is computationally expensive by increasing the number of iterative steps $K$, so both $K$ and mini-batch ($N$) size are upper-bound on the GPU memory. Therefore, to tackle this problem, we use the Truncated Backpropagation Through Time (TBPTT) algorithm as do in \cite{kokkinos2019iterative} to train our network, where the sequence is unrolled into a small number of $k$-steps out of total $K$ and then the back-propagation is performed on the small $k$-steps. Furthermore, we compute the $\ell_1$-Loss with respect to GT images after $k$ iterative steps according to Eq.~\eqref{eq:l1loss}. 
\begin{table}[htbp!]
\caption{The settings of input LR and corresponding HR patch sizes during training.}
    \vspace{-0.18cm}
	\centering
	\resizebox{.35\textwidth}{!}
	{\begin{tabular}{|c|c|c|}
			\hline
			Scale factor     &  LR Patch size &   HR Patch size     \\
			\hline
			$\times 2$       & $60\times60$ & $120\times120$ \\\hline
			$\times 3$       & $50\times50$ & $150\times150$\\\hline
			$\times 4$       & $40\times40$ & $160\times160$ \\\hline
	\end{tabular}}	
	\label{tab:input_size}
\end{table}

\section{Experiments}
\label{sec:exp}

\begin{table*}[htbp!]	
    \caption{Average PSNR/SSIM values for scale factors $\times 2$, $\times 3$, and $\times 4$ with bicubic degradation model. The best performance is shown in {\color{red} red} and the second best performance is shown in {\color{blue} blue}.}
    \vspace{-0.25cm}
	\centering	
	\resizebox{0.85\textwidth}{!}{\begin{tabular}{|c|c|c|c|c|c|c|c|c|c|}
			\hline
			\multirow{2}{*}{Dataset}  & \multirow{2}{*}{Scale} & \multirow{2}{*}{Bicubic} & SRCNN~\cite{dong2014srcnneccv} & VDSR~\cite{kim2016vdsrcvpr} & EDSR-baseline~\cite{Lim2017edsrcvprw}  &  RISR~\cite{jiang2018risr} & SRFBN-S~\cite{Li2019srfbncvpr} & ISRResCNet & ISRResCNet+\\
			&                        &                          &   (ECCV-2014)    &    (CVPR-2016)       &  (CVPR-2017)   &  (ICPR-2018)    &   (CVPR-2019)   &  (Ours) &  (Ours)  \\ \hline\hline
			\multirow{3}{*}{Set5}     
			& $\times2$ & 33.55 / 0.9304   & 36.16 / 0.9509    &  37.30 / 0.9573   &  37.59 / {\color{red}0.9605}     & 37.63 / 0.9590      & 37.39 / 0.9597  &  {\color{blue}37.67} / 0.9596    & {\color{red}37.79} / {\color{blue}0.9600} 	 \\
			& $\times3$ & 30.35 / 0.8686   & 32.28 / 0.9020   &  33.50 / 0.9197  &  {\color{blue}34.18} / {\color{red}0.9270}     & 33.91 / 0.9234     & 33.99 / 0.9252  &  34.08 / 0.9251    & {\color{red}34.20} / {\color{blue}0.9258}     \\  
			& $\times4$ & 28.39 / 0.8109   & 29.99 / 0.8519    &  31.20 / 0.8818  &  {\color{red}31.89} / {\color{red}0.8932}      & 31.58 / 0.8870      & 31.76 / {\color{blue}0.8914}  &  31.63 / 0.8890   & {\color{blue}31.77} / 0.8908 \\ \hline\hline
			\multirow{3}{*}{Set14}    
			& $\times2$ & 30.05 / 0.8701   & 31.81 / 0.9033    &  32.84 / 0.9121      &  {\color{red}33.21} / {\color{red}0.9177}   & {\color{blue}33.16} / 0.9133   & 33.04 / {\color{blue}0.9157}  &  32.89 / 0.9144   & 33.06 / 0.9155   \\
			& $\times3$ & 27.40 / 0.7763   & 28.70 / 0.8151    &  29.54 / 0.8323      &  {\color{red}29.91} / {\color{red}0.8421}    & {\color{red}29.91} / 0.8338   & 29.72 / 0.8376  &  29.63 / 0.8365    & {\color{blue}29.76} / {\color{blue}0.8381}   \\  
			& $\times4$ & 25.86 / 0.7056   & 26.92 / 0.7427    &  27.75 / 0.7688      &  {\color{red}28.20} / {\color{red}0.7820}    & {\color{blue}28.19} / 0.7707   & 28.05 / {\color{blue}0.7785}  &  27.99 / 0.7757    & 28.08 / 0.7776  \\ \hline\hline
			\multirow{3}{*}{B100}     
			& $\times2$ & 29.51 / 0.8439   & 31.07 / 0.8838    &  31.83 / 0.8949      &  {\color{red}32.03} / {\color{red}0.8996}  & {\color{blue}32.01} / 0.8968  & 31.87 / 0.8972  &  31.98 / 0.8974    & {\color{red}32.03} / {\color{blue}0.8980}      \\  
			& $\times3$ & 27.19 / 0.7399   & 28.17 / 0.7799    &  28.80 / 0.7971      &  {\color{red}29.03} / {\color{red}0.8056}   & 28.92 / 0.7996   & 28.90 / 0.8015  &  28.91 / 0.8014    & {\color{blue}28.96} / {\color{blue}0.8024}   \\  
			& $\times4$ & 25.96 / 0.6698   & 26.70 / 0.7029    &  27.27 / 0.7252      &  {\color{red}27.53} / {\color{red}0.7365}    & 27.37 / 0.7270   & 27.41 / {\color{blue}0.7321}  &  27.40 / 0.7301  & {\color{blue}27.44} / 0.7313      \\ \hline\hline
			\multirow{3}{*}{Urban100} 
			& $\times2$ & 26.84 / 0.8409   & 29.01 / 0.8885    &  30.67 / 0.9129      &  {\color{red}31.81} / {\color{red}0.9271}     & 31.06 / 0.9168    & 31.27 / 0.9208  &  31.29 / 0.9205    & {\color{blue}31.45} / {\color{blue}0.9220}     \\  
			& $\times3$ & 24.44 / 0.7359   & 25.82 / 0.7874    &  27.09 / 0.8271     &  {\color{red}28.05} / {\color{red}0.8524}     & 27.41 / 0.8338   & 27.60 / 0.8418  &  27.57 / 0.8409    & {\color{blue}27.70} / {\color{blue}0.8432}   \\  
			& $\times4$ & 23.13 / 0.6593   & 24.11 / 0.7051    &  25.14 / 0.7522      &  {\color{red}25.98} / {\color{red}0.7850}       & 25.41 / 0.7595    & {\color{blue}25.66} / {\color{blue}0.7725}  &  25.56 / 0.7682    & 25.65 / 0.7705       \\ \hline
	\end{tabular}}
	\label{tab:comp_sota_bi}
\end{table*}

\subsection{Data augmentation}
We use DIV2K\cite{div2k} dataset that contains 800 HR images for training. We take the input LR image patches as a bicubic downsample (\ie regarded as a standard degradation) with their corresponding HR image patches. We augment the training data with random vertical and horizontal flipping, and $90^{\circ}$ rotations. Moreover, we also consider another effective data augmentation technique, called \emph{MixUp} \cite{zhang2017mixup}. In \emph{Mixup}, we take randomly two samples $(\bx_i , \by_i)$ and $(\bx_j , \by_j)$ in the training HR/LR set $(\Tilde{\bX}, \bY)$ and then form a new sample $(\Tilde{\bx} , \by)$ by interpolation of the pair samples by following the same degradation model \eqref{eq:degradation_model} as do in \cite{feng2019suppressing}. This simple technique encourages our network to support linear behavior among training samples. 

\subsection{Technical details}
We use the RGB input LR and corresponding HR patches with different patch sizes according to the upscaling factor as listed in Table~\ref{tab:input_size}. We train the network for 300 epochs with a batch size of 4 using the Adam optimizer with parameters $\beta_1 =0.9$, $\beta_2=0.999$, and $\epsilon=10^{-8}$ without weight decay to minimize the $\ell_1$-Loss~\eqref{eq:l1loss}. We use the method of kaiming He~\cite{He2015DelvingDI} to initial the \emph{Conv} weights and bias to zero. The learning rate is initially set to $10^{-3}$ for the first 100 epochs and then multiplies by $0.5$ for every 50 epochs. We set the number of iterative steps ($K$) to $20$ and feedback steps (FB) to 4 for our method. The extrapolation weights $\mathbf{w}\in\mbR^K$ are initialized with $\mathbf{w}_t = \frac{t^k-1}{t^k+2}, \forall 1\le t \le K$, and then further fine-tune on the training data as do in\cite{kokkinos2019iterative}. The projection layer parameter $\sigma$ is estimated according to \cite{liu2013single} from the input LR image. We initialize the projection layer parameter $\alpha$ on a log-scale value from $\alpha_{max}=2$ to $\alpha_{min}=1$ and then further fine-tune during the training via back-propagation. In order to further enhance the performance of our network, we use a self-ensemble strategy~\cite{timofte2016seven} (denoted as ISRResCNet+), where the LR inputs are flipped/rotated and the SR results are aligned and averaged for enhanced prediction.

\subsection{Evaluation metrics and SR benchmarks}
We evaluate the trained model under the Peak Signal-to-Noise Ratio (PSNR) and Structural Similarity (SSIM) metrics on four benchmark datasets: Set5~\cite{set5}, Set14~\cite{set14}, B100~\cite{b100}, and Urban100~\cite{urban100}. In order to keep a fair comparison with existing networks, the quantitative SR results are only evaluated on $Y$ (luminance) channel of the transformed $YCbCr$ color space.

\begin{figure}[htbp!]
    \centering
    \begin{subfigure}[t]{0.5\textwidth}
        \centering
        \includegraphics[width=5.0cm]{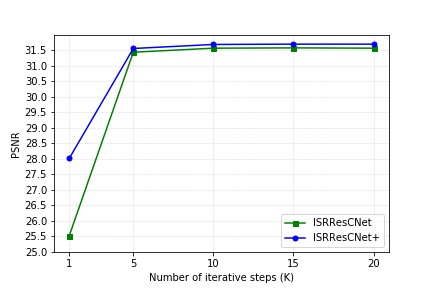}
        \caption{PSNR vs. K}
    \end{subfigure}\\
    \begin{subfigure}[t]{0.5\textwidth}
        \centering
        \includegraphics[width=5.0cm]{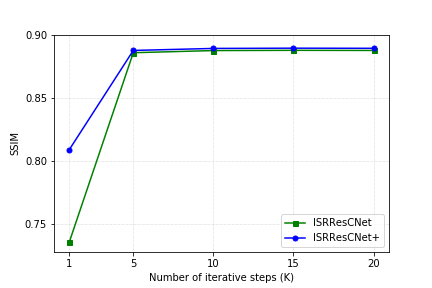}
        \caption{SSIM vs. K}
    \end{subfigure}
    \caption{Average PSNR/SSIM performance (Set5 on $\times4$) of proposed ISRResCNet and ISRResCNet+ after each iterative step (K).}
    \label{fig:psnr_ssim_vs_k}
\end{figure}
\begin{table}[htbp!]
\caption{The impact of iterative (K) and feedback (FB) steps on ISRResCNet on the scale factor $\times4$. The average PSNR/SSIM values are evaluated on Set5 testset.}
    \vspace{-0.4cm}
	\begin{center}
		\resizebox{.48\textwidth}{!}{\begin{tabular}{|c|c|c|c|c|c|c|}
				\hline
				 {\begin{tabular}[c]{@{}c@{}}~Feedback~ \\ steps~(FB)\end{tabular}} &
				 {\begin{tabular}[c]{@{}c@{}}~Iterative~ \\ steps~(K)\end{tabular}} &
				 {\begin{tabular}[c]{@{}c@{}}~\#Params~ \\ 
				 ($\times10^3$)\end{tabular}}  &  
				 {\begin{tabular}[c]{@{}c@{}}~ResBlocks~ \\ (D)\end{tabular}} & {\begin{tabular}[c]{@{}c@{}}~Feature-Maps~ \\ (F)\end{tabular}} & 
				 ISRResCNet & ISRResCNet+
				\\ \hline \hline
				None & 10 & 380  & 5 & 64 & 31.44 / 0.8855  & 31.59 / 0.8876   \\
				None & 20 & 380  & 5 & 64 & 31.56 / 0.8874   & 31.69 / 0.8891 \\
				4 & 10 & 388  & 5 & 64 & 31.63 / 0.8890   & 31.77 / 0.8908 \\
				\hline
		\end{tabular}}
		\smallskip
		\label{tab:ablation_table}
	\end{center}
\end{table}
\subsection{Ablation study of iterative (K) and feedback (FB) steps}
\label{sec:perf_tradeoff}
For our ablation study, we evaluate our proposed ISRResCNet and ISRResCNet+ performance on Set5 benchmark dataset at $\times4$ upscaling factor. Table~\ref{tab:ablation_table} shows the average PSNR/SSIM performance after iterative steps ($K$) and feedback (FB) steps. Our trained model achieves better performance (PSNR/SSIM) by increasing the number of iterative steps $1$ to $20$ with the shared network parameters (\ie 380K) without using \emph{FB} steps (see in Fig.~\ref{fig:psnr_ssim_vs_k} and Table~\ref{tab:ablation_table}). When the \emph{FB} connections introduce into our network, the model converges in the less number of iterative steps (\ie $10$) with better reconstruction results by requiring a few additional parameters (\ie +8K) because these error feedback connections~\cite{Li2019srfbncvpr} after residual blocks provide strong early reconstruction ability. Since these error feedbacks are beneficial on the higher scale ($\times4$), so we report the quantitative results in the Table~\ref{tab:comp_sota_bi} with feedback steps at $\times4$ upscaling factor, while the others ($\times2, \times3$) are without feedback steps with $20$ iterative steps. It can also be noted (see Fig.~\ref{fig:psnr_ssim_vs_k}) that a few iterative steps (e.g. $5$) are enough to obtain excellent SR results with the performance trade-off between quantitative results and the computation time of our method.
\begin{figure}[htbp!]
    \centering
    \begin{subfigure}[t]{0.5\textwidth}
        \centering
        \includegraphics[width=8.5cm]{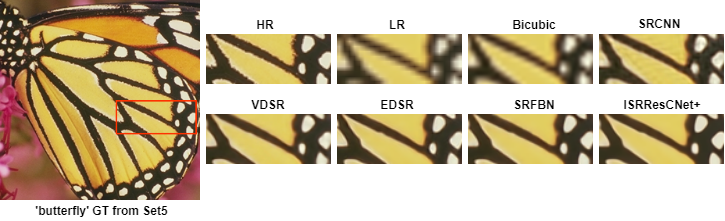}
    \end{subfigure}\\ 
    \begin{subfigure}[t]{0.5\textwidth}
        \centering
        \includegraphics[width=8.5cm]{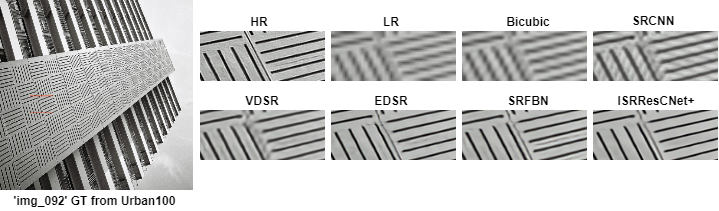}
    \end{subfigure}
    \caption{Visual comparison of our method with other state-of-art methods on $\times4$ super-resolution.}
    \label{fig:4x_result}
\end{figure}
\subsection{Comparison with the state-of-art methods}
We compare our method with other state-of-art SISR methods including SRCNN~\cite{dong2014srcnneccv}, VDSR~\cite{kim2016vdsrcvpr}, EDSR~\cite{Lim2017edsrcvprw}, RISR~\cite{jiang2018risr}, and SRFBN~\cite{Li2019srfbncvpr}, whose source codes are available online except for RISR method for which the quantitative results are directly taken from the paper. We run all the source codes by default parameters test settings through all the experiments. We report the quantitative results of our method with others in the Table~\ref{tab:comp_sota_bi}.  Our method exhibits better improvement in PSNR and SSIM compared to other methods, except the EDSR. Since the EDSR has a much deeper network containing 16 residual blocks with $1.5M$ parameters, while our model contains 5 residual blocks with $380K$ parameters, which is a much lighter model than EDSR with slightly performance difference in the PSNR ( \ie $+0.12dB$ on Set5) at $\times 4$ upscaling factor. Despite that, the parameters of the proposed network are much less than the other state-of-art SISR networks, which makes it suitable for deployment in mobile devices where memory storage and cpu power are limited as well as good image reconstruction quality (see section~\ref{sec:perf_tradeoff}).

Regarding the visual quality, Fig.~\ref{fig:4x_result} shows the visual comparison of our method with other SR methods for a high ($\times 4$) upscaling factor. The proposed method successfully reconstructs the good textures regions, sharp edges, and finer details of SR image compared to the other methods. 

\section{Conclusion}
We proposed a deep iterative residual CNNs for a single image super-resolution task by following the image observation (physical / real-world settings) model. The proposed method solves the SISR problem in an iterative manner by minimizing the discriminative loss function with residual learning. Our model requires few trainable parameters in comparison to other competing methods. The proposed network exploits  powerful image regularization and large-scale optimization techniques for image restoration. Our method achieves excellent SR results in terms of PSNR/SSIM and visual quality by following the real-world settings for limited memory storage and cpu power requirements for the mobile/embedded deployment.





%

\bibliographystyle{IEEEtran}
\bibliography{refs}

\end{document}